# Thermodynamic holography

Bo-Bo Wei[1,], Zhan-Feng Jiang[1,] & Ren-Bao Liu[1,2]

1. *Department of Physics and Centre for Quantum Coherence, The Chinese University of Hong Kong, Hong Kong, China*
2. *Institute of Theoretical Physics, The Chinese University of Hong Kong, Hong Kong, China*

**The holographic principle states that the information about a volume of a system is encoded on the boundary surface of the volume. Holography appears in many branches of physics, such as optics, electromagnetism, many-body physics, quantum gravity, and string theory. Here we show that holography is also an underlying principle in thermodynamics, a most important foundation of physics. The thermodynamics of a system is fully determined by its partition function. We prove that the partition function of a physical system is an analytic function of all the physical parameters, and therefore its values in any area on the complex plane of a physical parameter are uniquely determined by its values along the boundary. The thermodynamic holography has applications in studying thermodynamics of nano-scale systems (such as molecule engines, nano-generators and macromolecules) and provides a new approach to many-body physics.**

# Introduction

The most famous example of holography is probably the optical hologram, where the three-dimensional view of an object is recorded in a two-dimensional graph[1]. The holographic principle indeed has profound implications in many branches of physics. In electromagnetism, for instance, the electrostatic potential in a volume is uniquely determined by its values at the surface boundary[2]. Density functional theory[3,4], which is the foundation of quantum chemistry and first-principle calculations[4], may be viewed as a holography that maps the full ground-state wave function of a many-electron system (a complex function in an enormously high-dimensional space) to the ground state single-particle density (a real function in three-dimensional space). The holographic principle has also been shown relevant in quantum gravity where the quantum fields may be described by a theory in a lower dimension[5], and in string theory where the description of a volume of space can be encoded on the boundary of the region[6,7]. The holographic principle may also provide a novel approach to tackling strongly-correlated systems in condensed matter physics[8].

The physical properties of a system at thermodynamic equilibrium are fully determined by the partition function $\Xi(\beta, \lambda_1, \lambda_2, ..., \lambda_K)$ as a function of coupling parameters $\{\lambda_k\}$ and the temperature $T$ (or the inverse temperature $\beta \equiv 1/T$). The partition function is the summation of the Boltzmann factor $e^{-\beta H(\lambda_1, \lambda_2, ..., \lambda_K)}$ over all energy eigen states, i.e., $\Xi = \text{Tr}\left[e^{-\beta H(\lambda_1, \lambda_2, ..., \lambda_K)}\right]$, where the Hamiltonian $H = \sum_k \lambda_k H_k$ is characterized by a set of coupling parameters $\{\lambda_k\}$ (e.g., in spin models the magnetic field $h = \lambda_1$, the nearest neighbor coupling $J = \lambda_2$, and the next nearest neighbor coupling $J' = \lambda_3$, etc.). The normalized Boltzmann factor $\Xi^{-1} \exp(-\beta H)$ is the probability of the system in a state with energy $H$. In the following we consider one of the physical parameters of the system, $\lambda \in \{\lambda_1, \lambda_2, ..., \lambda_K\}$ and suppress the other parameters for the simplicity of notation. We assume that the Hamiltonian is a linear function of the parameter, but the main results (the theorems and the corollaries) in this paper automatically apply to Hamiltonians that are general analytic functions of the parameters.

In this article, we establish that the holographic principle holds in thermodynamics, for an arbitrary physical system. We prove that the partition function of a physical system is an analytic function of all the physical parameters, and, according to Cauchy theorem[9], its values in any area on the complex plane of a physical parameter are uniquely determined by its values along the boundary. Since the partition function with a complex parameter is equivalent to the coherence of a quantum probe[10, 11, 12], it is physically feasible to determine the whole thermodynamic properties of a system by measuring the probe spin coherence for just one value of the parameter. We theoretically studied an experimentally realizable system, namely, a nitrogen-vacancy centre coupled to a mechanical resonator, to demonstrate that the free energy of the resonator is fully determined by the nitrogen-vacancy centre spin decoherence for just one setting of parameters.

## Results

**Holography of partition function**

The thermodynamic holography is that the partition function of a physical system in an area of the complex plane of a physical parameter is uniquely determined by its values along the boundary. This stems from the Cauchy theorem in complex analysis [9] for analytic functions. The analyticity of partition functions is based on the following Theorem (see Fig. 1**a** for illustration):

*Theorem 1.* If the partition function $\Xi(\lambda_R)$ of a physical system is bounded for a real parameter $\lambda_R \in (\lambda_{\min}, \lambda_{\max})$, then the partition function continued to the complex plane of the parameter $\Xi(\lambda_R + i\lambda_I)$ is analytic where $\lambda_R \in (\lambda_{\min}, \lambda_{\max})$.

The proof of the theorem is given in Methods. Here by "physical system" we mean the basis states of the system are discrete, which is always possible by quantizing the system in a sufficiently large box. For finite (but arbitrarily large) systems of spins and fermions on lattices, the partition function is analytic in the whole complex plane of physical parameters. For bosons, the partition function can be non-analytical in certain regions (e.g., the partition functions of free bosons have singularities along the imaginary axis of frequencies, and for coupled bosons the system become unstable when the coupling is too strong), but as long as the partition function exists in certain ranges of real parameters, it is always analytic for arbitrary imaginary parts added to the parameters. This essentially proves that the partition functions of all physical systems are analytic when the parameters are extended vertically in the complex plane.

Here comes the main result of this paper. According to Cauchy theorem and Theorem 1, the partition function satisfies

$$\Xi(\beta,\lambda') = \frac{1}{2\pi i}\oint_C \frac{\Xi(\beta,\lambda)}{\lambda-\lambda'}d\lambda, \tag{1}$$

where $\lambda'$ is a complex parameter enclosed by the integration contour $C$ in the complex plane of $\lambda$ where the partition function is analytic (see Fig. 1**b**). A convenient choice of the boundary (integration contour) can be a rectangular path (see Fig. 1**c**) that consists of two straight lines perpendicular to the real axis, whose real parts $\lambda_\Re$ are respectively $\lambda_1$ and $\lambda_2$, and two segments parallel to the real axis, whose imaginary parts are respectively $-\infty$ and $+\infty$. The contributions from the two segments at infinity vanish (see Methods for details). Thus we have the following theorem:

*Theorem 2.* The partition function along two vertical lines in the complex plane of a physical parameter where the partition function is analytic uniquely determines the partition function at the any complex parameter between these two vertical lines, i.e.,

$$\Xi(\beta,\lambda') = \int_{-\infty}^{\infty}\frac{\Xi(\beta,\lambda_2+i\lambda_I)}{\lambda_2+i\lambda_I-\lambda'}\frac{d\lambda_I}{2\pi} - \int_{-\infty}^{\infty}\frac{\Xi(\beta,\lambda_1+i\lambda_I)}{\lambda_1+i\lambda_I-\lambda'}\frac{d\lambda_I}{2\pi}, \tag{2}$$

for $\lambda_1 < \Re(\lambda') < \lambda_2$.

Theorem 2 can be simplified by introducing a constant $M_-$ less than the minimum eigenvalue of $\partial_\lambda H$ to ensure $e^{\beta\lambda M_-}\Xi(\beta,\lambda)$ vanishes as $\Re(\lambda)\to+\infty$. That leads to

*Corollary 1.* If the partition function is analytic on the half complex plane to the right of a vertical line in the complex plane of a physical parameter, the partition function along this line uniquely determines the partition function on the half complex plane, that is,

$$\Xi(\beta,\lambda') = -\int_{-\infty}^{\infty} \frac{e^{\beta(\lambda_1+i\lambda_1-\lambda')M_-}\Xi(\beta,\lambda_1+i\lambda_1)}{\lambda_1+i\lambda_1-\lambda'} \frac{d\lambda_1}{2\pi}, \quad (3)$$

for $\lambda_1 < \Re(\lambda') < \infty$.

Similarly, by introducing a constant $M_+$ greater than the maximum eigenvalue of $\partial_\lambda H$ to ensure $e^{\beta\lambda M_+}\Xi(\beta,\lambda)$ vanishes as $\Re(\lambda) \to -\infty$, we have

*Corollary 2.* If the partition function is analytic on the half complex plane to the left of a vertical line in the complex plane of a physical parameter, the partition function along this line uniquely determines the partition function on the half complex plane, that is,

$$\Xi(\beta,\lambda') = \int_{-\infty}^{\infty} \frac{e^{\beta(\lambda_2+i\lambda_1-\lambda')M_+}\Xi(\beta,\lambda_2+i\lambda_1)}{\lambda_2+i\lambda_1-\lambda'} \frac{d\lambda_1}{2\pi}, \quad (4)$$

for $-\infty < \Re(\lambda') < \lambda_2$.

Corollaries 1 and 2 are particularly usefully for bosons, whose partition functions may be analytic only on half complex planes of parameters.

**Possible experimental realization of thermodynamic holography**

The thermodynamic holography is experimentally realizable. This is possible due to a recent discovery that the partition function of a system with a complex parameter is equivalent to the quantum coherence of a probe spin coupled to the system [10, 11, 12]. One can use a probe spin-1/2 ($S_z \equiv |\uparrow\rangle\langle\uparrow| - |\downarrow\rangle\langle\downarrow|$) coupled to a system (bath) with Hamiltonian, $H(\lambda)$, and the probe-bath interaction $H' = -\eta S_z \otimes H_I$, where $\eta$ is a small coupling constant and $H_I = \partial_\lambda H$. If the probe spin is initialized in the superposition state $|\uparrow\rangle + |\downarrow\rangle$ and the system in the thermal equilibrium with density matrix $\Xi^{-1}\exp(-\beta H)$, the quantum coherence of the probe spin, quantified by the spin polarization $\langle S_x(t)\rangle + i\langle S_y(t)\rangle = \text{Tr}\left[e^{-\beta H}e^{i(H+H')t}|\uparrow\rangle\langle\downarrow|e^{-i(H+H')t}\right]/\Xi(\beta,\lambda)$, has an intriguing form as [10, 11]

$$\langle S_+(\lambda,t)\rangle \equiv \langle S_x(\lambda,t)\rangle + i\langle S_y(\lambda,t)\rangle = \frac{\Xi(\beta,\lambda+it\eta/\beta)}{\Xi(\beta,\lambda)}. \quad (5)$$

The quantum coherence of the probe spin is equivalent to the partition function with a complex parameter, $\lambda + it\eta/\beta$ and the evolution time serves as the imaginary part of the physical parameter. Recently, Lee-Yang zeros have been observed via such a measurement [12]. In terms of the probe spin coherence, equation (2) can be rewritten as

$$\Xi(\beta,\lambda') = \Xi(\beta,\lambda_2)\int_{-\infty}^{\infty}\frac{\langle S_+(\lambda_2,t)\rangle}{\lambda_2 + i\eta t/\beta - \lambda'}\frac{\eta dt}{2\pi\beta} - \Xi(\beta,\lambda_1)\int_{-\infty}^{\infty}\frac{\langle S_+(\lambda_1,t)\rangle}{\lambda_1 + i\eta t/\beta - \lambda'}\frac{\eta dt}{2\pi\beta}. \quad (6)$$

Similarly, equation (3) can be rewritten as

$$\Xi(\beta,\lambda') = -\Xi(\beta,\lambda_1)\int_{-\infty}^{\infty}\frac{e^{\beta(\lambda_1+i\eta t/\beta-\lambda')M_-}\langle S_+(\lambda_1,t)\rangle}{\lambda_1 + i\eta t/\beta - \lambda'}\frac{\eta dt}{2\pi\beta}, \quad (7)$$

and equation (4) as

$$\Xi(\beta,\lambda') = \Xi(\beta,\lambda_2)\int_{-\infty}^{\infty}\frac{e^{\beta(\lambda_2+i\eta t/\beta-\lambda')M_+}\langle S_+(\lambda_2,t)\rangle}{\lambda_2 + i\eta t/\beta - \lambda'}\frac{\eta dt}{2\pi\beta}. \quad (8)$$

Note that the probe spin coherence resembles the form of quantum quenches[17]. Therefore the quantum quench dynamics may also be studied using the thermodynamic holography.

Equations (6)-(8) establish an experimentally implementable holographic approach to thermodynamics. Let $\lambda'$ be an arbitrary real number, we can determine the free energy $F = -\beta^{-1}\ln(\Xi)$ by

$$e^{-\beta[F(\lambda')-F(\lambda)]} = \text{sgn}(\lambda-\lambda')\int_{-\infty}^{\infty}\frac{e^{\beta(\lambda+i\eta t/\beta-\lambda')M_{\text{sgn}(\lambda-\lambda')}}\langle S_+(\lambda,t)\rangle}{\lambda + i\eta t/\beta - \lambda'}\frac{\eta dt}{2\pi\beta}. \quad (9)$$

Thus we can extract the full thermodynamic properties of the system from probe spin coherence measurement for just a single value of the physical parameter. Note that previously the free energy difference ($\Delta F$) has been related to the work ($\Delta W$) in a non-equilibrium physical process by the Jarzynski equality $\exp(-\beta\Delta F) = \langle\exp(-\beta\Delta W)\rangle$ [18]. The Jarzynski equality is particularly useful for determining free energy differences for small thermodynamic systems such as quantum engines and biomolecular systems[16, 19, 20]. Using the thermodynamic holography and the Jarzynski equality, we establish a general relation between the probe spin coherence, the work done on small systems, and the free energy changes. This general relation is indeed the foundation of two recent proposals for experimental measurement of the characteristic function of the work distributions [21, 22] in quantum quenches[17], which plays a central role in the fluctuation relations in non-equilibrium thermodynamics [23]. The power of the thermodynamic holography is that one can obtain free energy change for any parameters using the probe spin coherence measurement for just one value of the parameter instead of quenching the system to various parameters[16, 19, 20].

The thermodynamic holography can also be used to determine the probe spin coherence for an arbitrary parameter by the coherence measurement for just one value of the parameter. Choosing a complex parameter $\lambda' + it'\eta/\beta$, we have the probe spin coherence

$$\langle S_+(\lambda',t')\rangle = \frac{\int_{-\infty}^{\infty} dt \frac{e^{\beta(\lambda+i\eta t/\beta-\lambda'-i\eta t'/\beta)M_{\text{sgn}(\lambda-\lambda')}}\langle S_+(\lambda,t)\rangle}{\lambda+i\eta t/\beta-\lambda'-i\eta t'/\beta}}{\int_{-\infty}^{\infty} \frac{e^{\beta(\lambda+i\eta t/\beta-\lambda')M_{\text{sgn}(\lambda-\lambda')}}\langle S_+(\lambda,t)\rangle}{\lambda+i\eta t/\beta-\lambda'}dt}. \qquad (10)$$

## Discussion

**Thermodynamic holography of a mechanical resonator coupled to a probe spin**

To demonstrate the idea of thermodynamic holography, we study an experimentally realizable system as the model example, namely, a nitrogen-vacancy (NV) centre spin coupled to a nano-mechanical resonator [24, 25] (see Fig. 2**a**). This model may also be realized in a superconducting resonator [26]. The NV centre has a spin triplet ground state ($S=1$) with a large zero field splitting $\Delta = 2.87\,\text{GHz}$. A mechanical resonator with frequency $\omega \sim 3\,\text{GHz}$ has a magnetic tip. The NV centre is placed right under the tip. The oscillation of the mechanical resonator generates a time dependent magnetic field on the NV centre spin with an interaction Hamiltonian, $V = \delta(a^+ + a)\otimes S_x$. Under realistic conditions, the coupling $\delta$ can reach 1 MHz for a magnetic tip with size of 100 nm and an NV centre located about 25 nm under the tip [27]. The Hamiltonian of coupled mechanical resonator and the NV center is $H = \Delta S_z^2 + \omega a^+ a + \delta S_x(a^+ + a)$. We make use of the spin states $|+1\rangle$ and $|0\rangle$ as a probe and define $\sigma_z \equiv |+1\rangle\langle+1| - |0\rangle\langle0|$ and the corresponding spin flip operators $\sigma^+ \equiv |+1\rangle\langle0|$ and $\sigma^- \equiv |0\rangle\langle+1|$. Since the coupling $\delta(\leq 1\,\text{MHz}) \ll \omega(\sim 3\,\text{GHz})\,\&\,\Delta(\sim 2.87\,\text{GHz})$, the perturbation theory (see Methods for details) gives an effective Hamiltonian

$$H_{\text{eff}} \approx \Delta\sigma_z + \omega a^+ a + \eta\sigma_z \otimes a^+ a, \quad (11)$$

where $\eta = 4\delta^2\omega/(\Delta^2 - \omega^2)$. The NV center acts as a probe spin and the mechanical resonator as a system (bath). The partition function $\Xi(\beta,\omega) = (1-e^{-\beta\omega})^{-1}$ of the oscillator has an infinite number of singularity points $\omega = i2n\pi/\beta$ (where $n$ is an arbitrary integer) along the imaginary axis of the frequency. But the partition function is finite for all positive frequency. Therefore, according to Theorem 1, the partition function is analytic in the half complex plane of frequency with positive real part. Hence the holographic principle applies. We shall demonstrate that the free energy difference can be extracted from the probe spin coherence measurement of the NV center. We assume that the NV center is initialized in the superposition state $|1\rangle + |0\rangle$ and the mechanical resonator in the thermal equilibrium. Note that in the current case the probe spin coherence is a periodic function of time since the energy levels of the oscillator are equally spaced. So the probe spin coherence in one period of time, from 0 to $2\pi/\eta$, is sufficient to yield the full information of the partition function. The measurement time of the NV center spin coherence is $\sim 2\pi/\eta \approx 100\ \mu s$, which is within the spin coherence time of an NV centre in an isotopically purified diamond [28].

Figure 2**b** presents the real (red solid line) and imaginary (blue dashed line) parts of the NV centre spin coherence as functions of time for the resonator at temperature 150 mK ($\beta\omega$=1). From the spin coherence we can obtain the free energy (relative to the value at $\beta\omega$=1) for arbitrary $\beta\omega$. Fig. 2**c** shows the exponentiated free energy difference (blue dots) constructed from the NV center spin coherence (Fig. 2**b**). The result obtained by the holographic method agrees well with the direct calculation of the free energy (the solid red line in Fig. 2**c**).

The holographic approach can also be used to determine the probe spin coherence for arbitrary $\beta\omega$. The real and imaginary parts of the constructed spin coherence for $\beta\omega$=2 as functions of time are presented respectively in Fig. 3**a** & 3**b**, which agree well with the direct solution.

## Summary

In this work we have established the concept of thermodynamic holography, which states that the partition function of a physical system in an area of the complex plane of a physical parameter is uniquely determined by its values along the boundary. Since the partition function with a complex parameter is equivalent to the probe spin coherence, one can experimentally implement the thermodynamic holography through probe spin coherence measurement for just one physical parameter. Thermodynamic holography may have applications in studying thermodynamics of nano-scale systems (such as molecule engines[13], nano-generators [14] and macromolecules [15, 16]) and provide a new approach to many-body physics.

## Methods

**Proof of Theorem 1.**

We first consider the case that the system has a finite number ($K$) of basis states. By Taylor expansion, we define a series of functions $\Xi^{(N)}(\lambda) = \sum_{n=0}^{N} \frac{1}{n!} \text{Tr}\left[-\beta H(\lambda)\right]^n$, which are finite sums of polynomial functions of $\lambda$. Obviously, all these functions are analytic in the whole complex plane of $\lambda$. Considering the parameter in the range $|\lambda| < \Lambda$, for an arbitrarily small quantity $\varepsilon$, there exists an integer $N_\varepsilon$ such that $\sum_{n=N_\varepsilon+1}^{\infty} \frac{h^n}{n!} < \varepsilon$, where $h$ is the maximum of $\left|\text{Tr}\left[-\beta H(\lambda)\right]\right|$ for $|\lambda| < \Lambda$. Therefore $\left|\Xi^{(N)}(\lambda) - \Xi(\lambda)\right| < \varepsilon$ for $N > N_\varepsilon$, i.e., the function series uniformly converge to the partition function in the parameter range $|\lambda| < \Lambda$. According to uniform convergence theorem of analytic functions [9], the partition function is analytic for $|\lambda| < \Lambda$. Since $\Lambda$ can be chosen arbitrarily, we get the following lemma:

*Lemma* 1. The partition function of a quantum system that has a finite number of basis state is analytic for any finite parameter $\lambda$ in the whole complex plane.

Now we consider the general cases. Since we can always confine a physical system using a sufficiently large box, we can assume that the basis states of the system are discrete (which, however, can be infinite). We denote the discrete basis states as $\{|0\rangle,|1\rangle,|2\rangle,...\}$ and define truncated partition functions $\tilde{\Xi}_K(\lambda) = \sum_{k=0}^{K} \langle k|e^{-\beta H(\lambda)}|k\rangle$ in a finite subspace $\{|0\rangle,|1\rangle,|2\rangle,...,|K\rangle\}$. According to Lemma 1, the truncated partition functions are all analytic. If the partition function is bounded ($\Xi(\lambda_R) \leq \Xi_{max}$) for a real parameter $\lambda_R \in (\lambda_{min}, \lambda_{max})$, we can choose the basis states such that $\tilde{\Xi}_K(\lambda_R)$ uniformly converges to $\Xi(\lambda_R)$ as $K \to \infty$, i.e., for an arbitrarily small quantity $\varepsilon$, there exists an integer $K_\varepsilon$ such that $|\tilde{\Xi}_K(\lambda_R) - \Xi(\lambda_R)| < \varepsilon$ for $K > K_\varepsilon$, or, $\sum_{k>K_\varepsilon} \langle k|e^{-\beta H(\lambda_R)}|k\rangle < \varepsilon$..

By Hölder's inequality for finite-dimensional matrices [29], $|\text{Tr}[A^\dagger B]| \leq \sqrt{\text{Tr}[A^\dagger A]}\sqrt{\text{Tr}[B^\dagger B]}$, we have $|\langle k|e^{-\beta H+i\lambda_I H_I}|k\rangle| \leq \langle k|e^{-\beta H/2+i\lambda_I H_I/2}e^{-\beta H/2-i\lambda_I H_I/2}|k\rangle$ (the Hölder's inequality in the one-dimensional subspace expanded by the basis state $|k\rangle$). In addition, according to the Bernstein inequality [30], $\text{Tr}[e^{A^\dagger}e^A] \leq \text{Tr}[e^{A^\dagger+A}]$ for any finite-dimensional matrix $A$, we have the following lemma:

*Lemma* 2. For any basis state $|k\rangle$, real number $\lambda_I$ and Hermitian Hamiltonians $H$ and $H_I$, $|\langle k|e^{-\beta H+i\lambda_I H_I}|k\rangle| \leq \langle k|e^{-\beta H}|k\rangle$. Furthermore, in any Hilbert space, $|\text{Tr}(e^{-\beta H+i\lambda_I H_I})| \leq \text{Tr}(e^{-\beta H})$.

Lemma 2 means the partition function at the real axis of a parameter is always greater than its vertical continuation on the complex plane. According to Lemma 2, $\left|\sum_{k>K_\varepsilon} \langle k|e^{-\beta H(\lambda_R)-i\beta\lambda_I H_I}|k\rangle\right| < \sum_{k>K_\varepsilon} \langle k|e^{-\beta H(\lambda_R)}|k\rangle < \varepsilon$. So the series of analytic functions $\tilde{\Xi}_K(\lambda_R+i\lambda_I)$ uniformly converge to $\Xi(\lambda_R+i\lambda_I)$ for $\lambda_R \in (\lambda_{min}, \lambda_{max})$ as $K \to \infty$. According to uniform convergence theorem of analytic functions [9], the partition function $\Xi(\lambda_R+i\lambda_I)$ is analytic for $\lambda_R \in (\lambda_{min}, \lambda_{max})$.

That proves Theorem 1.

**Proof of Theorem 2.**

We just need to show $\int_{\lambda_1}^{\lambda_2} \frac{\Xi(\beta, \lambda_R + i\lambda_I)}{\lambda_R + i\lambda_I - \lambda'} \frac{d\lambda_R}{2\pi} \to 0$ for $|\lambda_I| \to \infty$. Suffices it to show $\Xi(\beta, \lambda_R + i\lambda_I) = \text{Tr}\left[ e^{-\beta H - i\lambda_I H_I} \right]$ is bounded for $\lambda_R \in (\lambda_{\min}, \lambda_{\max})$. According to Lemma 2, $|\Xi(\beta, \lambda_R + i\lambda_I)| \leq \Xi(\beta, \lambda_R)$. Since $\Xi(\lambda_R) \leq \Xi_{\max}$ for $\lambda_R \in (\lambda_{\min}, \lambda_{\max})$, $\Xi(\beta, \lambda_R + i\lambda_I)$ is bounded. Therefore Theorem 2 is proved.

**Derivation of the effective Hamiltonian of a coupled spin-oscillator system.**

The coupling Hamiltonian for the pseudo spin and the oscillator is $H = \Delta\sigma_z + \omega a^+ a + \delta\sigma_x(a^+ + a)$. Since $\delta \ll \omega, \Delta$, we can treat $\delta\sigma_x(a^+ + a)$ as a perturbation. A unitary transformation defined by $U = \exp\left[ 2\delta\omega / (\Delta^2 - \omega^2)(\sigma^+ a - \sigma^- a^+) \right]$ leads to $UHU^+ \approx \Delta\sigma_z + \omega a^+ a + \left[ 4\delta^2\omega / (\Delta^2 - \omega^2) \right]\sigma_z \otimes a^+ a + \left[ \delta^2\omega / (\Delta^2 - \omega^2) \right]\sigma_z$. The last term can be dropped since it is only a small correction to the Zeeman energy of the spin. We therefore obtain equation (11).

**Numerical method.**

In determining the free energies of the mechanical oscillator from probe spin coherence of the NV centre through thermodynamic holography, we have uniformly taken 100 data points of the NV centre spin coherence within one period ($[0, 2\pi/\eta]$) and then carried out numerical integrations, which result in the data for the blue symbols in Figs. 2**c**, 3**a** & 3**b**.

## References

1. Hecht, E. & Zajac, A. *Optics* (Addison-Wesley, 1974).
2. Griffiths, D. J. *Introduction to Electrodynamics, 4${}^{th}$ Ed.* (Addison-Wesley, 2012).

3. Hohenberg, P. & Kohn, W. Inhomogenous Electron Gas. *Phys. Rev.* **136**, B864 (1964).

4. Kohn, W. & Sham, L. J. Self-consistent equations including exchange and correlation effects. *Phys. Rev.* **140**, A1133 (1965).

5. Stephens, C. R., 't Hooft, G. & Whiting, B. F. Black hole evaporation without information loss. *Class. Quantum Grav.* **11,** 621 (1994).

6. Susskind, L. The World as a Hologram. *J. Math. Phys.* **36**, 6377 (1995).

7. Witten, E. Anti-de Sitter space and holography. *Adv. Theor. Math. Phys*. **2**, 253 (1998).

8. Witczak-Krempa, W., Sørensen, E. S. & Sachdev, S. The dynamics of quantum criticality revealed by quantum Monte Carlo and holography, *Nature Phys.* **10**, 361-366 (2014).

9. Gamelin, T. W. *Complex Analysis*. (Springer-Verlag, New York, 2001).

10. Wei, B. B., & Liu, R. B. Lee-Yang Zeros and Critical Times in Decoherence of a Probe Spin Coupled to a Bath. *Phys. Rev. Lett.* **109**, 185701 (2012).

11. Wei, B. B., Chen, S. W., Po, H. C. & Liu, R. B. Phase transitions in the complex plane of physical parameters. *Sci. Rep.* **4**, 5202 (2014).

12. Peng, X. H., Zhou, H., Wei, B. B., Cui, J. Y., Du, J. F. & Liu, R. B. Observation of Lee-Yang zeros. *arxiv:1403. 5383* (2014).

13. Bockrath, M. W. A Single-Molecule engine, *Science* **338**, 754-755 (2012).

14. Wang, Z. L. & Song, J. Piezoelectric Nanogenerators Based on Zinc Oxide Nanowire Arrays, *Science* **312**, 242–246 (2006).

15. Gratzer, W. *Giant Molecules: From nylon to nanotubes*. (Oxford University Press, 2011).

16. Collin, D., Ritort, F., Jarzynski, C., Smith, S. B., Tinoco, I. & Bustamante, C. Verification of the Crooks fluctuation theorem and recovery of RNA folding free energies. *Nature* **437**, 231-234 (2005).

17. Polkovnikov A., Sengupta K., Silva A., & Vengalattore M. *Colloquium:* Nonequilibrium dynamics of closed interacting quantum systems. *Rev. Mod. Phys*. **83**, 863 (2011).

18. Jarzynski, C. Nonequilibrium equality for free energy differenece. *Phys. Rev. Lett.* **78**, 2690 (1997).

19. Hummer, G. & Szabo, A., Free energy reconstruction from nonequilibrium single-molecule pulling experiments, *Proc. Natl. Acad. Sci.* **98**, 3658 (2002).

20. Liphardt, J., Dumont, S., Smith, S. B., Tinoco, I.Jr. & Bustamante, C. Equilibrium information from nonequilibrium measurements in an experimental test of Jarzynski's equality. *Science* **296**, 1832–1835 (2002).

21. Dorner R., Clark S. R., Heaney L., Fazio R., Goold J. & Vedral V. Extracting Quantum Work Statistics and Fluctuation Theorems by Single-Qubit Interferometry. *Phys. Rev. Lett.* **110**,230601 (2013).

22. Mazzola L., Chiara G. D. & M. Paternostro M. Measuring the Characteristic Function of the Work Distribution. *Phys. Rev. Lett.* **110**,230602 (2013).

23. Campisi M., Hanggi P. & Talkner P. *Colloquium*: Quantum fluctuation relations: Foundations and applications. *Rev. Mod. Phys.* **83**, 771 (2011).

24. Rabl, P. *et al*. Strong magnetic coupling between an electronic spin qubit and a mechanical oscillator. *Phys. Rev. B* **79**, 041302 (2009).

25. Kolkowitz, S. *et al.* Coherent sensing of a mechanical resonator with a single-spin qubit. *Science* **335**, 1063–1606 (2012).

26. LaHaye, M. D., Suh, J., Echternach, P. M., Schwab, K. C. & Roukes, M. L. Nanomechanical measurements of a superconducting qubit. *Nature* **459**, 960–964 (2009).

27. Mamin, H. J., Poggio, M., Degen, C. L. & Rugar, D. Nuclear magnetic resonance imaging with 90-nm resolution. *Nature Nanotechnol*. **2**, 301–306 (2007).

28. Maurer, P. C. *et al.* Room-temperature quantum bit memory exceeding one second. *Science* **336**, 1283–1286 (2012).

29. Horn, R. A. & Johnson, C. R. *Matrix Analysis, 2$^{nd}$ Ed*. (Cambridge University Press, 2012).

30. Bernstein, D. S. Inequalities for the trace of matrix exponentials. *SIAM J. Matrix Anal. Appl.* **9**, 156 (1988).

**Acknowledgements** This work was supported by Hong Kong Research Grants Council/General Research Fund CUHK402410, The Chinese University of Hong Kong Focused Investments Scheme, and Hong Kong Research Grants Council/Collaborative Research Fund HKU8/CRF/11G.

**Author Contributions** R.B.L. conceived the idea, designed the project and formulated the theory. B.B.W. & R.B.L. designed the model. B.B.W. and R.B.L. carried out the research. Z.F.J. participated in the study in its initial stage. R.B.L. & B.B.W. wrote the paper.

**Author Information** The authors declare no competing financial interests. Correspondence and requests for materials should be addressed to R.B.L. (rbliu@phy.cuhk.edu.hk).

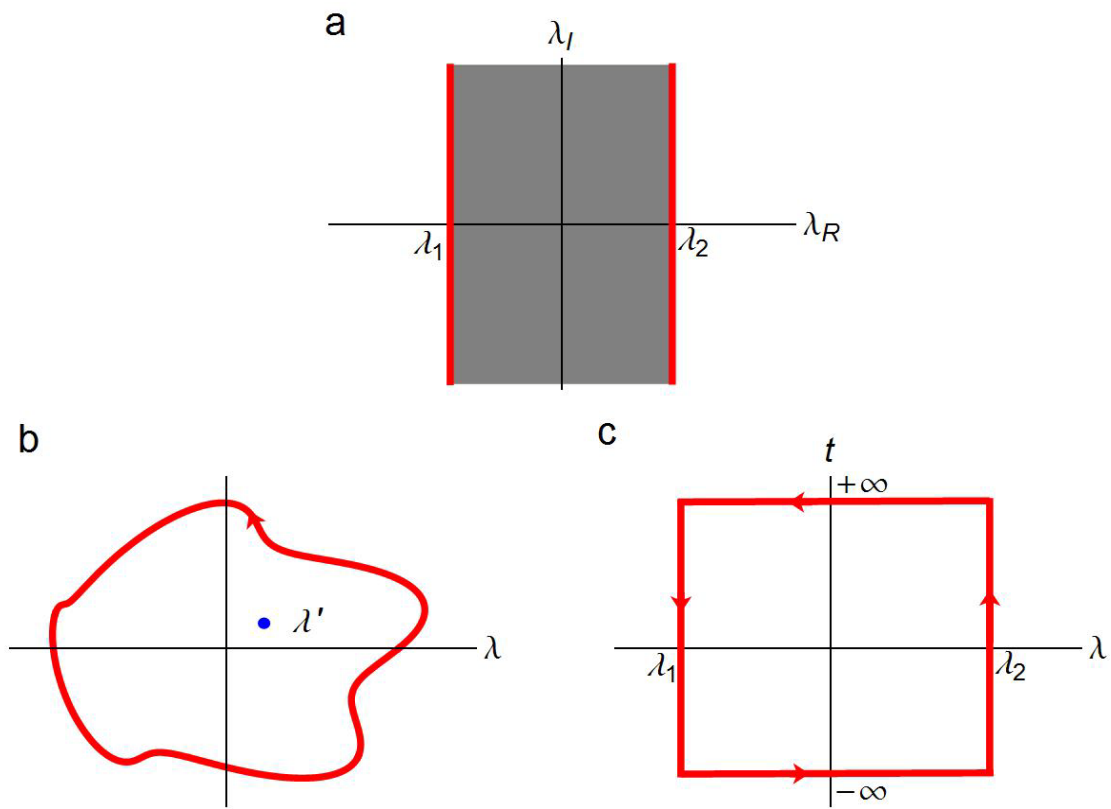

**Figure 1 | Holography of partition function in the complex plane of a physical parameter. a**, If the partition function is bounded in a segment $(\lambda_1, \lambda_2)$ on the real axis of the parameter, it is analytical on the region (shadowed) on complex plane extended vertically from the segment, $\{\lambda_R + i\lambda_I | \lambda_1 < \lambda_R < \lambda_2\}$. **b**. Integration contour in the complex plane of $\lambda$. The red solid curve is the integration contour and $\lambda'$ is a point inside the analytic domain. **c**. A rectangular integration contour in the complex plane of a physical parameter $\lambda$ with $t$ being the imaginary part. Two vertical straight lines are parallel to the $t$-axis, with real parts being $\lambda_1$ and $\lambda_2$, and imaginary parts extended from $-\infty$ to $+\infty$. The other two segments parallel to the real axis with imaginary parts at $-\infty$ or $+\infty$.

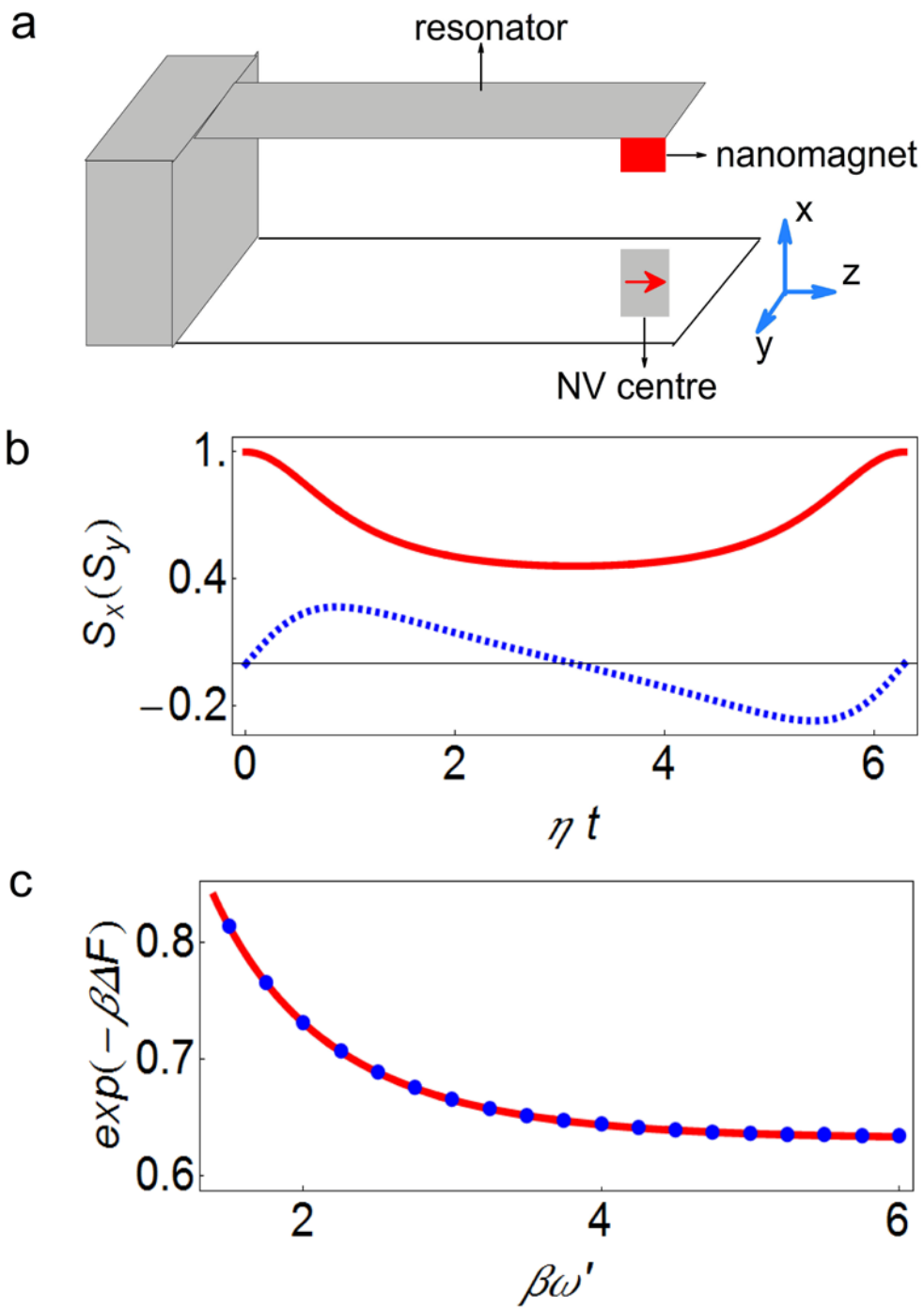

**Figure 2 | Extracting free energy difference of an oscillator from the probe spin coherence measurement. a.** Schematic plot of an NV centre coupled to a mechanical resonator. The mechanical resonator has a magnetic tip attached at the end. An NV centre is placed right under the magnetic tip. The oscillation of the mechanical resonator generates a time-dependent magnetic field on   the

NV centre spin. **b**. The spin coherence of the NV center as a function of scaled time, $\eta t$ with $\eta = 4\delta^2 \omega / (\Delta^2 - \omega^2)$, at temperature $\beta\omega$=1. The red solid line presents $S_x$ and the dashed blue line shows $S_y$ .**c**. The free energy difference for the mechanical resonator obtained from spin coherence of the NV center. The red solid line is the direct solution and the blue dot is the result of holography using integration of the spin coherence of the NV center at $\beta\omega$=1.

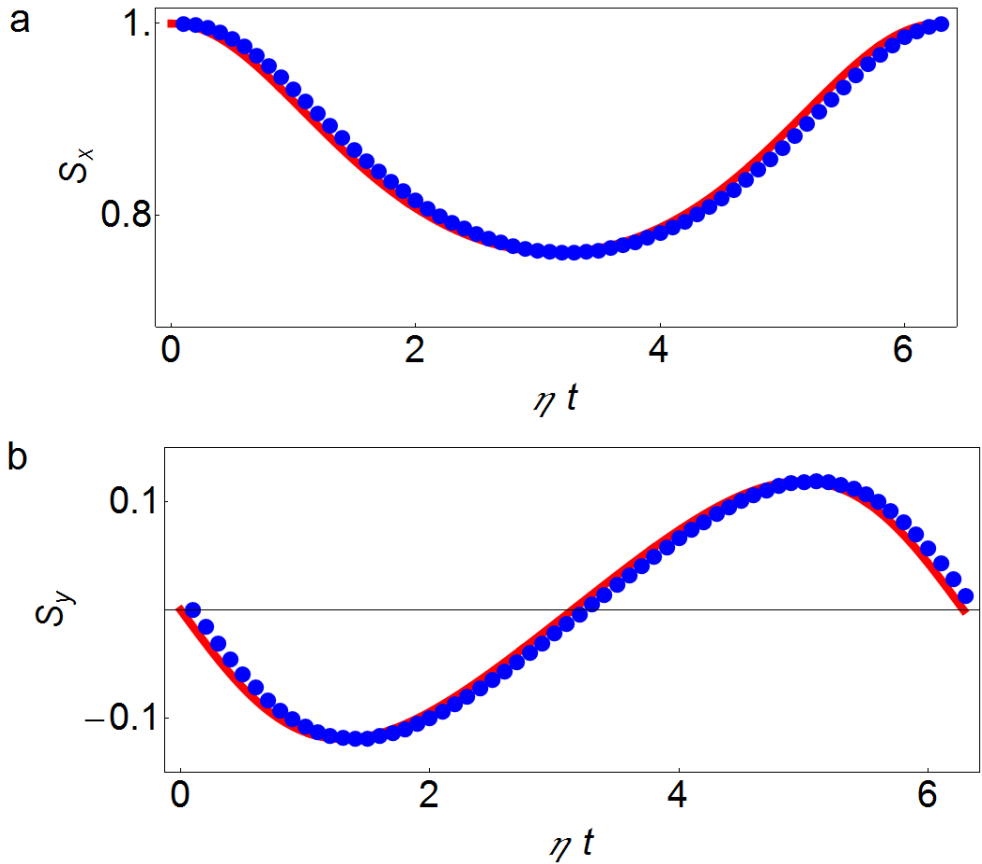

**Figure 3 | Probe spin coherence by thermodynamic holography. a.** The real part of spin coherence of the NV centre at $\beta\omega=2$ as a function of scaled time, $\eta t$ with $\eta = 4\delta^2\omega/(\Delta^2-\omega^2)$. The blue dots are obtained by integration of the spin coherence at $\beta\omega=1$, and the solid red line is the direct solution. **b.** The same as **a** but for the imaginary part of the spin coherence.